# Anisotropic $Bi_2O_2Se(Te)$ Monolayer: Realizing Ultra-High Carrier Mobility and Giant Electric Polarization in Two-Dimension

*Jiewen Xiao[†, §], Yuxin Wang[†, §], Xiang Feng[†, §], Dominik Legut[‡], Tianshuai Wang[†], Yanchen Fan[†], Tonghui Su[†], Rushi Gong[†], Ruifeng Zhang[†] & Qianfan Zhang*[†]*

[†]*School of Materials Science and Engineering, Beihang University, Beijing 100191, P. R. China.*

[‡]*IT4Innovations & Nanotechnology Centre, VSB-Technical University of Ostrava, 17.listopadu 15, Ostrava CZ-70833, Czech Republic.*

[§]**Author contributions:** *These authors contributed equally to this work.*

**Corresponding authors:** qianfan@buaa.edu.cn.

**Abstract**

Here, we have identified the monolayer phase of $Bi_2O_2Se$ as a promising two-dimensional semiconductor with ultra-high carrier mobility and giant electric polarization. Due to the strong reconstruction originated from the interlayer electrostatic force, we have applied structure prediction algorithms to explore the crystalline geometry of $Bi_2O_2Se$ monolayer with the lowest total energy. Considering Se and Te belong to the same group, $Bi_2O_2Te$ monolayer is also investigated based on a similar scheme. Further calculations suggest that the high carrier mobility is maintained in the monolayer phase and the moderate band gap will lead to the strong optical absorption in the visible light region. In particular, the electron mobility in $Bi_2O_2Te$ can reach as high as 3610 $cm^2V^{-1}s^{-1}$ at room temperature, which is almost ten times of conventional transition metal dichalcogenides (TMD) family. Because of the strong structural anisotropy, a remarkable spontaneous in-plane and out-of-plane electric polarization is additionally revealed along with significant piezoelectric properties, endowing them as promising candidates in the area of photovoltaic solar cells, optoelectronic materials and field effect transistors.

## Introduction

Since the discovery of graphene[1] in 2004, 2D layered materials have attracted considerable attention, and more and more novel 2D materials have been theoretically predicted and experimentally fabricated, for instance black phosphorene[2], $In_2Se_3$[3] and the transition metal dichalcogenides (TMD) family[4]. Due to their attractive physical and chemical properties in low-dimension, they have promising applications in the field of catalysis, sensors and nano-size electronic devices. However, various 2D layered materials suffer from their own problems and are still far from practical applications. For instance, graphene possesses excellent stability along with remarkable carrier mobility, but suffers from the zero-band gap, thus being unable to serve as semiconducting switch[5]. $MoS_2$, as a typical member of TMD family, exhibits a desirable direct band gap (~1.8 eV), whose potential application in the field of optoelectronic devices is further limited by the low carrier mobility[6]. While the new discovered black phosphorene, despite possessing attractive transportation properties, faces the degradation problem when being exposed in air or humid environments[7]. On the other hand, intrinsic 2D materials are usually with centrosymmetric structures while in-plane and especially out-of-plane anisotropy is still rare, although this endows the possibility of spontaneous electric polarization and further provides a great opportunity to develop novel devices in the area of field-effect transistors (FET). Therefore, the theoretical exploration on new 2D semiconductors remains as a critical topic.

Recently, layered materials $Bi_2O_2Se$ has received great attention because of its ultra-high mobility, robust band gap and excellent stability when exposed in the air[8][9]. Therefore, those desirable properties make $Bi_2O_2Se$ being considered as one of the most promising next-generation semiconductors and its potential usage in the field of optoelectronics and ferroelectrics has also been revealed[10][11]. Meanwhile, $Bi_2O_2S$ and $Bi_2O_2Te$, with the similar crystalline structure and element composition, are also widely studied[11]. To realize its application as nano-size electronic devices, achieving the ultra-thin phase of $Bi_2O_2Se$, is the key topic in both the experiments and theoretical calculations. However, compared to the van der Waals force in conventional layered materials, the interlayer electrostatic interaction in the bulk phase $Bi_2O_2Se$ is rather strong, which will lead to the dynamic instability and thus the remarkable change of the structure when being reduced to the monolayer. Besides, such variance of geometric structures is promising to further induce a substantial change on their semiconducting and optical properties. Nevertheless, due to the experimental limitations, the atomic configuration of monolayer phase is still not identified and its physical properties in low-dimension remain unclear.

In the present work, we theoretically predicted the energetically favored $Bi_2O_2Se(Te)$ monolayer based on the structure prediction algorithm. It is found that the monolayer phase undergoes a considerable reconstruction and exhibits strong in-plane and out-of-plane anisotropy. Further calculation on electronic structures shows that the $Bi_2O_2Se(Te)$ monolayer not only maintains the superior transport properties with carrier mobility as high as 3610 $cm^2V^{-1}s^{-1}$, but also possesses excellent optical absorption property that is better than intrinsic silicon. Particularly, the remarkable reconstruction further endows $Bi_2O_2Se(Te)$ monolayer the giant spontaneous electric polarization, which opens up the great opportunity for their further applications as nano-size electronic devices.

**Computational method**

The projector augmented wave (PAW)[12][13] formalism of density functional theory (DFT) as implemented in the Vienna *Ab-initio* Simulation Package (VASP)[14][15] is used for the total energy and electronic structure calculations. The Gaussian smearing method was adopted with the smearing width as 0.02 eV. The energy cutoff for plane-wave expansion of the PAWs is set to 600 eV. Brillouin zone was sampled by the Gamma-centered Monkhorst-Pack method with 8×12×1 grid[16]. In the vertical direction, a vacuum layer of about 20 Å in thickness is introduced for all the two-dimensional layered structure, which is thick enough to avoid the artificial interaction in the *z* direction. To simulate electronic structures, we applied both traditional GGA-PBE method[17] and HSE06 (Heyd-Scuseria-Ernzerhof) hybrid functional[18][19]. Former method is very efficient and can be used for most simulations, while the latter one can result in reasonable band gap, which is very important for the computation on the optical properties. To rationally treating the heavy atoms such as Bi and Se(Te), we also considered the spin-orbit coupling (SOC) interaction for the electronic structure simulation.

The energetically favorite geometry of two-dimensional monolayered $Bi_2O_2Se(Te)$ (M-$Bi_2O_2Se(Te)$) is explored by the crystalline structure prediction computation scheme. Herein, we applied the structure prediction on two-dimensional phases by the Crystal Structure Analysis by Particle Swarm Optimization (CALYPSO)[20][21]. Such method only requires chemical compositions for a given compound to predict stable or metastable structures at given external conditions (e.g., pressure and temperature). To prove the dynamical stability of the predicted structures, we use PHONOPY[22] to simulate the lattice dynamics (phonon dispersion and phonon density of states), while the *ab-initio* molecular dynamics (AIMD)[23][24] scheme is also performed in order to further verify their thermal stabilities at room temperature.

## Results and discussion

### 1. Structure and stability

The crystalline structure of $Bi_2O_2Se$ and $Bi_2O_2Te$ monolayer is firstly investigated starting from their bulk counterparts. Figure S1 exhibits geometric structures of bulk phases along with the hypothesized monolayered $Bi_2O_2Se$ (M-$Bi_2O_2Se$), being consistent with conventional view of the structures of 2D materials based on their layered bulk phases. To determine the phase stability, the atomic vibrations (lattice dynamic) simulation is applied[25]. Here the phonon dispersion spectra exhibits considerable imaginary frequencies (Figure S2). Such imaginary vibration modes are related to the absence of strong interlayer electrostatic force in bulk phases, and suggest the considerable reconstruction of the hypothesized monolayer. In addition, similar layered structure for M-$Bi_2O_2Te$ is also dynamically unstable, as reflected from the phonon spectra shown in Figure S2.

To account for different kinds of potential reconstruction, structure prediction algorithm is applied and the number of formula units per unit cell is set to be 1, 2, 3, 4, 6, corresponding to 5, 10, 15, 20, 30 atoms. Among more than 25,000 crystalline structures, we identified the one with the lowest total energy. Both M-$Bi_2O_2Se$ and M-$Bi_2O_2Te$ possess the similar crystalline structure, as expected, and the atomic configuration and detailed geometric information are shown in Figure 1a and in Table S1-S3 respectively. Compared to the layered unit in the bulk phase, M-$Bi_2O_2Se(Te)$ is with an enlarged formula unit $Bi_4O_4Se(Te)_2$ per unit cell, and the symmetry belongs to the space group of *Pm* (space group No. 6). Reconstructed configuration of M-$Bi_2O_2Se(Te)$ contains two kinds of layers: the Janus Se(Te)-Bi-O monolayer and the Bi-O layer and from bottom to top, which are connected closely by Bi-O bonds. On the side of Bi-O-Se layer, atoms are hexagonally packed with elongated Bi-O bonds ranging from 2.58 Å to 2.68 Å, while the top Bi-O layer shows the parallelogram channel in the *y* direction with shorter Bi-O bonds around 2.20 Å. Due to the distinct structure in *x* and *y* direction for the Bi-O layer above, the hexagonal patterns for Se-Bi-O atomic layer lying below distort accordingly, and the three originally symmetric Bi-Se/Te bonds are unequal to each other. Thus, this structure varies greatly compared to the layered unit in bulk phase and exhibits the desirable in-plane and out-of-plane anisotropic geometry, which is very critical for the formation of spontaneous electric polarization as shown in the following text.

To demonstrate their dynamic stability, phonon spectra of discovered monolayer phase is calculated and shown in Figure 1b and S4. Very tiny imaginary frequency exists at Γ point for M-$Bi_2O_2Se$ since the supercell used to simulate lattice dynamics is with limited size, which is constrained by computational consumptions. While there are completely no imaginary modes for M-$Bi_2O_2Te$, demonstrating their excellent dynamical stabilities compare to bulk counterparts. *Ab-initio* molecular dynamics simulation (AIMD) on the 4×4×1 super-lattice of M-$Bi_2O_2Se$(Te) was additionally applied, while the simulated temperature is set to 300 K. In Figure S5, we exhibited the snapshots of atomic configurations of M-$Bi_2O_2Se$(Te) at 0 *ps*, 5 *ps* and 10 *ps* during AIMD simulation. The parallelogram structure of Bi-O atomic layer stacked by the Se-Bi-O hexagonal layer can be well retained. Besides, during the whole AIMD simulation time, the almost invariant free energies of M-$Bi_2O_2Se$ and M-$Bi_2O_2Te$ (Figure 1b and S4) also confirm their thermodynamic stabilities at room temperature.

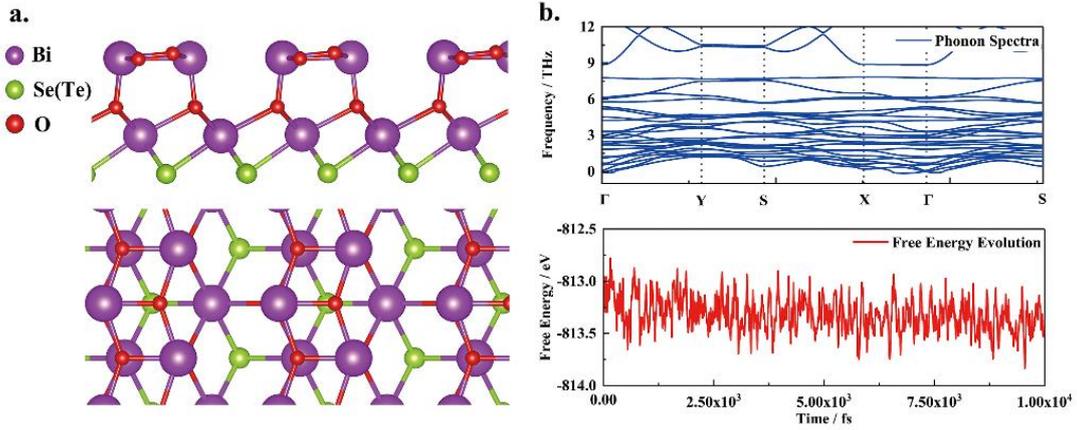

**Figure 1.** a. Side and top views of discovered M-Bi$_2$O$_2$Se(Te), where purple, green and red spheres represent Bi, Se (Te) and O atoms. b. Phonon spectra (upper panel) and free energy evolution during AIMD simulation (lower panel) for M-Bi$_2$O$_2$Se.

## 2. Electronic structure and transport properties

The band structure of M-Bi$_2$O$_2$Se(Te) was further calculated based on HSE06+SOC scheme in order to achieve more accurate band gap and include SOC effect. As shown in Figure 2a and 2b, M-Bi$_2$O$_2$Se is an indirect semiconductor with band gap as 1.77 eV while M-Bi$_2$O$_2$Te possesses the desirable direct band gap with a moderate value as 1.24 eV. Additionally, the band gap of bulk phase Bi$_2$O$_2$Se was simulated (Figure S6), being consistent with the detected value in experiments[8]. It is worth to note that the inclusion of SOC is essential, which greatly affects the edge states of valance band in M-Bi$_2$O$_2$Se and conduction band in M-Bi$_2$O$_2$Te. For M-Bi$_2$O$_2$Se, the inclusion of SOC will lead to a dispersive peak in the Γ-X direction, thus altering the original position of valance band maximum (VBM) (Figure S7). While for M-Bi$_2$O$_2$Te, compared to the HSE06 result in Figure S7, SOC moves the position of conduction band maximum (CBM) from X to Γ point and further lowers its energy, endowing M-Bi$_2$O$_2$Te as a semiconductor with moderate direct band gap.

Partial densities of states (PDOS) are also simulated as presented in the right panel of Figure 2a and 2b. Results indicate that electrons in the conduction band are mainly composed of empty *p* states of Bi since Bi atoms possess the valance state as +3 in the monolayer. Detailed analysis shows that there are two kinds of Bi atoms. $Bi_1$ is from the above Bi-O parallelogram channel, bonding with O atom and possessing negligible states right below Fermi level. $Bi_2$ sits in the Se-Bi-O Janus monolayer, whose ionic Bi-O bond leads to the empty conduction states. Meanwhile, there is an additional Bi-Se/Te bonding state appeared below Fermi level, which is mainly contributed from *p* orbits of Se(Te) and further reflects the major differences among Se and Te compounds. For M-$Bi_2O_2$Se, states below Fermi level are quite sharp and localized, while the states for telluride compound are flatter and the PDOS is reduced to zero more gradually, as denoted by the arrow in Figure 2a and 2b. Since the electronegativities for Bi (2.02) and Te (2.10) are quite close to each other compared to the value 2.55 for Se, this can be understood as the ionic feature of Bi-Se bond while the Bi-Te bond is more covalent. Distinct distribution feature of electronic states can also be confirmed by the flat valance band for Se and the dispersive one for Te. Thus, it can be further predicted that less ionic bonds in M-$Bi_2O_2$Te will endow it the better the electric conductivity.

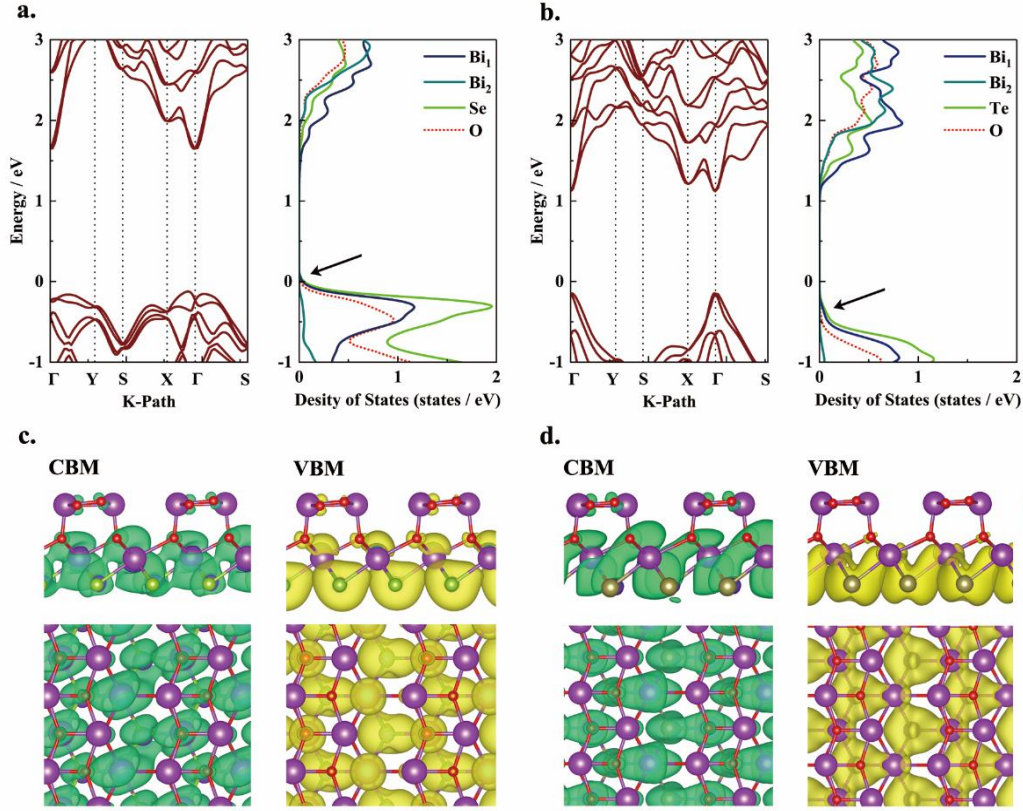

**Figure 2.** Band structure and partial density of states based on HSE06+SOC scheme for a. M-$Bi_2O_2Se$ and b. M-$Bi_2O_2Te$, where Γ (0, 0, 0), Y (0, 0.5, 0), S (0.5, 0.5, 0) and X (0.5, 0, 0) denote high symmetric points in the Brillouin zone. Side and top views of charge density spatial distribution (CDSD) for c. CBM and VBM of M-$Bi_2O_2Se$ and d. CBM and VBM of M-$Bi_2O_2Te$. The CBM is represented by the green iso-surface while the VBM is shown in yellow, where the iso-surface value is set to 0.0005 e/Å$^3$.

Considering the high carrier mobility in bulk phase $Bi_2O_2Se$, transport properties in low-dimension as M-$Bi_2O_2Se$(Te) were explored. Since the electron scattering caused by phonons is the main factor that constrains mobility, the phonon-limited scattering model here, which includes anisotropic features of effective mass, deformation potential and elastic modulus, and can be expressed as[26]:

$$\mu_{\alpha x} = \frac{e\hbar^3 (\frac{5C_{x\_2D} + 3C_{y\_2D}}{8})}{k_B T m^{*~3/2}_{\alpha x} m^{*~1/2}_{\alpha y} (\frac{9E^2_{\alpha x} + 7E_{\alpha x}E_{\alpha y} + 4E^2_{\alpha y}}{20})}$$

Here, $\alpha = e$ or $h$ (electrons or holes), denotes the type of carriers. $C_{x\_2D}$ and $C_{y\_2D}$ are 2D elastic stiffness coefficients along the $x$ and $y$ direction, which is scaled by the $z$ parameter corresponding to the spacing between 2D layers, and can be expressed as:

$$C_{\beta\_2D} = \frac{2(E - E_0)}{[S_0 (\Delta L_\beta / L_{\beta 0})^2]}$$

$\beta$ denotes lattice directions ($x$ and $y$). $E$ and $E_0$ are the total energies with and without changes ($\Delta L_\beta$) in the lattice constant $L_{\beta 0}$ while $S_0$ is the cross section of the 2D materials at the equilibrium. $m^*_{\alpha x}$ and $m^*_{\alpha y}$ are the effective mass of carrier $\alpha$ along the x and y directions, which are calculated by relation as: $m^*_\alpha = \hbar^2 [\partial^2 E(k) / \partial k^2]^{-1}$. While the deformation potential of carries in $x$ and $y$ direction are expressed as $E_{\alpha x}$ and $E_{\alpha y}$, which is defined based on $E_{\alpha,\beta} = \Delta E_{\alpha,\beta} / (\Delta L_\beta / L_{\beta 0})$. Where $\Delta E_{\alpha,\beta}$ is the energy change of the band edges of carrier $\alpha$ when dilatation or compression strain ($\Delta L_\beta / L_{\beta 0}$) is applied in the $\beta$ direction. The temperature T in the formulation is set to 300 K and interchanging $x$ with $y$ can yield $\mu_{\alpha y}$ accordingly.

Calculated results are shown in Table 1. It is worth to note that, a methodical literature survey has been performed and there is another less accurate formulation that has also been widely used[27], which still produces consistent results (section IV in supporting information). Table 1 shows that, at room temperature, M-Bi$_2$O$_2$Se possess carrier mobility ~ 1500 cm$^2$V$^{-1}$s$^{-1}$ while the value for M-Bi$_2$O$_2$Te is almost two times of that of M-Bi$_2$O$_2$Se. Compared to other 2D systems, the magnitude of carrier mobility in M-Bi$_2$O$_2$Se(Te) is quite remarkable. For instance, the carrier mobility of MX$_2$ family ( M = Mo, W; X = S, Se, Te) are mainly around 100 cm$^2$V$^{-1}$s$^{-1}$, and only WS$_2$ can barely reach as high as 1000 cm$^2$V$^{-1}$s$^{-1}$ at room temperature[28-30]. For another recently reported system as Group IVB dichalcogenide monolayers[31], most of candidates in this family possess carrier mobility below 600 cm$^2$V$^{-1}$s$^{-1}$. Furthermore, except the considerable magnitude of carrier mobility, the strong anisotropy is also reflected in M-Bi$_2$O$_2$Se, which exhibits that the value for the *x* direction is almost 10 times of the value in the *y* direction. While for M-Bi$_2$O$_2$Te, the carrier mobility is less anisotropic and the electrons in the *y* direction exhibit the most superior transport properties with the mobility as high as 3610 cm$^2$V$^{-1}$s$^{-1}$.

To understand the magnitude and variance of carrier mobility, there are two decisive factors: effective mass and the deformation potential. Here, we plotted the charge density spatial distribution (CDSD) for both CBM and VBM states to illustrate their effects, as shown in Figure 2c and 2d. It can be clearly seen that all edge states exhibit off-site feature, and a large component of hybridized states concentrates on the intermediate region apart from on-site atoms. Such broad spread of wavefunction can induce strong dispersion in momentum space, thus leading to small effective mass (0.1~0.3 $m_e$). This delocalized feature is more significant in M-Bi$_2$O$_2$Te and there is even a connected channel formed in the *y* direction for the VBM state. It is also worth to note that, for Bi$_2$O$_2$Se, the effective hole mass exhibits strong anisotropic feature, since CDSD shows the $p_x$ and $p_z$ hybridization for VBM that locates among the Γ-X line. However, valance band edges of M-Bi$_2$O$_2$Te on Γ point are composed of $p_x$ and $p_y$ hybridized states, endowing them isotropic in-plane delocalization and hence isotropic effective mass. On the other aspect, the carrier mobility is strongly affected by the deformation potential, which describes the dependence of band edges on applied strains. Results in Table 1 indicate that the response of electrons in CBM is generally more subtle compared to holes determined by VBM. It can be understood as that the more localized VBM concentrating on Se/Te atoms will lead to the less sensitivity against strain. While CBM is composed of delocalized bonding states among Bi-Se/Te, whose atomic distance will be directly affected by the applied strain. Therefore, variances on Bi-Se/Te bond further shift/lower conduction edges, leading to large deformation potential for electrons accordingly.

|  | $C_{x\_2D}$ (N/m) | $C_{y\_2D}$ (N/m) | Carrier Type | $m_x^*$ ($m_e$) | $m_y^*$ ($m_e$) | $E_x$ (eV) | $E_y$ (eV) | $\mu_x$ (cm$^2$V$^{-1}$s$^{-1}$) | $\mu_y$ (cm$^2$V$^{-1}$s$^{-1}$) |
|---|---|---|---|---|---|---|---|---|---|
| Bi$_2$O$_2$Se | 30.69 | 69.73 | E | 0.224 | 0.239 | 2.248 | 5.876 | 1352 | 100 |
|  |  |  | H | 0.153 | 1.107 | 2.268 | 3.912 | 1811 | 23 |
| Bi$_2$O$_2$Te | 27.03 | 70.75 | E | 0.136 | 0.127 | 3.144 | 5.092 | 3402 | 3610 |
|  |  |  | H | 0.221 | 0.228 | 2.680 | 2.734 | 2562 | 3077 |

**Table 1.** Calculated elastic modulus ($C_{x\_2D}$ and $C_{y\_2D}$), effective mass ($m_x^*$ and $m_y^*$), deformation potential ($E_x$ and $E_y$) and carrier mobility ($\mu_x$ and $\mu_y$) at 300 K.

## 3. Optical properties

Since the band gap of M-Bi$_2$O$_2$Se(Te) is moderate, it further suggests their potential applications in the field of photovoltaic and optoelectronic devices. Based on electronic structures from HSE06 scheme, we investigated the optical properties, and focused on the computation on the absorption coefficient. The imaginary part of dielectric function can be calculated based on the following expression[32]:

$$\varepsilon_{\alpha\beta}(\omega) = \frac{4\pi^2 e^2}{\Omega} \lim_{q \to 0} \frac{1}{q^2} \sum_{c,v,k} 2\omega_k \delta(\varepsilon_{ck} - \varepsilon_{vk} - \omega) \times \langle \mu_{ck+e_\alpha} | \mu_{vk} \rangle \langle \mu_{ck+e_\alpha} | \mu_{vk} \rangle^*$$

Where $\Omega$ is the volume of the unit cell and index $c$, $v$ and $k$ refers to the conduction band states, valance band states and wave vector. While $e_\alpha$ represents the unit vector for three Cartesian directions. $u_k$ is an eigenstate and $\omega_k$ denotes the weight of k-point. For the real part as $\varepsilon_1(\omega)$, it can be further obtained by the Kramers-Kronig transformation. Thus, the absorption coefficient $\alpha(\omega)$ can be achieved through the following formula[33][34]:

$$\alpha(\omega)=\sqrt{2}\omega\{\sqrt{\varepsilon_1^2(w)+\varepsilon_{\alpha\beta}^2(w)}-\varepsilon_1(w)\}^{1/2}$$

The evolution of $\alpha_{xx}$ and $\alpha_{yy}$ against wavelength are presented in Figure 3. It demonstrates that the in-plane absorption properties for both M-$Bi_2O_2Se$ and M-$Bi_2O_2Te$ are even better than that in the intrinsic silicon. Due to the anisotropic in-plane structure, the absorption coefficients in two directions are quite different. The absorption intensity in $y$ direction shows the advantage in short wavelength region (~ 490 nm for M-$Bi_2O_2Se$ and ~ 590 nm for M-$Bi_2O_2Te$), while the absorbance in $x$ direction is with the absorption peak in long wavelength region (~ 580 nm for M-$Bi_2O_2Se$ and ~ 710 nm for M-$Bi_2O_2Te$). M-$Bi_2O_2Te$ exhibits much wider absorption spectrum compared to M-$Bi_2O_2Se$. Particularly, the absorption intensity in the long wavelength region can reach as high as ~$10^5$ cm$^{-1}$ for M-$Bi_2O_2Te$, which is one order of magnitude larger than that of the intrinsic silicon. Therefore, due to the geometry reconstruction and the quantum confinement effect, moderate band gap of M-$Bi_2O_2Se(Te)$ phase can lead to excellent optical properties in both the visible light region and near infrared region. Therefore, both of them can be viewed as the promising layered materials for the potential applications in photovoltaic solar cells and optoelectronic devices, especially for M-$Bi_2O_2Te$.

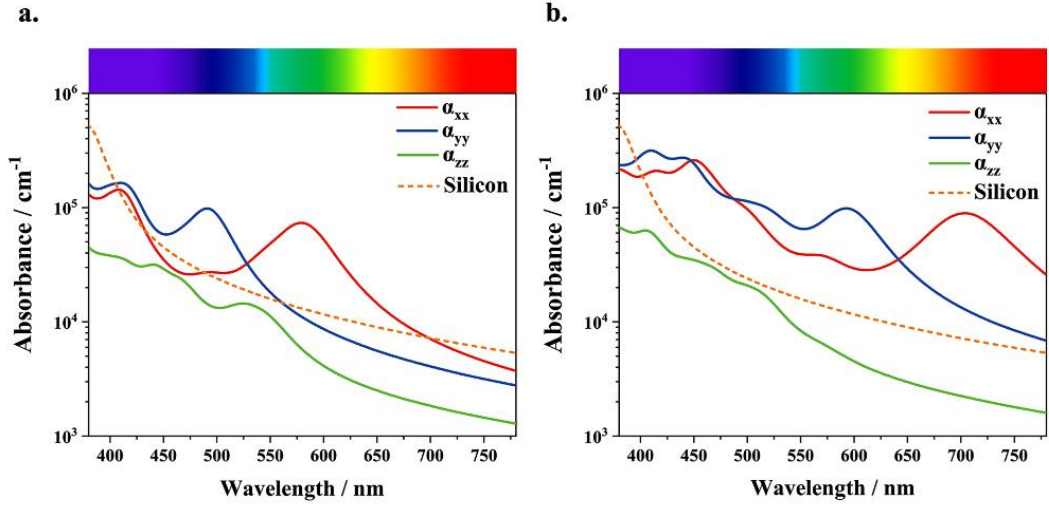

**Figure 3.** The evolution of tensor components of photo-absorption coefficient $\alpha$ against wavelength for a. M-$Bi_2O_2Se$ and b. M-$Bi_2O_2Te$, respectively. While the orange dash line represents the isotropic photo-absorption coefficient for the intrinsic silicon.

## 4. Giant spontaneous electric polarization and piezoelectric properties

Semiconducting and optical properties already exhibit anisotropic features that originate from the reconstructed geometries, and we further applied Berry phase scheme[35] to investigate the potential spontaneous electric polarization. Results indicate that the electric moment in *z* direction is 0.42 eÅ and 0.34 eÅ per unit cell for M-$Bi_2O_2Se$ and M-$Bi_2O_2Te$, which is one of the largest vertical polarization strengths according to our knowledge, and 3 - 4 times larger than that in the layered $In_2Se_3$ (~ 0.1 eÅ)[3]. Such giant vertical dipole moment is induced by the polar structure of M-$Bi_2O_2Se$(Te). To demonstrate this point, the atomic differential charge density of M-$Bi_2O_2Se$ is applied, which is defined as:

$$\Delta\rho = \rho_0 - \rho_a$$

$\rho_0$ is the charge density that is converged in an electronic self-consistent calculation and contains the bonding information among atoms, while $\rho_a$ is from a non-self-consistent calculation only summing over atomic charge densities. As plotted in Figure 4a, the spatial charge accumulation and depletion exhibits two polarized double layers with the same polarization direction as +$z$. Due to the great ability of O atom to capture electrons, the strong dipole polarization can be built in +$z$ direction among the upper Bi-O double layers. While the similar electron migration also happens in the below Bi-Se layers, further enhancing the polarization strength in +$z$ direction. Besides, the larger electronegativity for selenium than that for telluride lead to that fact that M-$Bi_2O_2Se$ exhibits stronger vertical polarization strength. On the other hand, the mirror symmetry is retained in the $y$ direction, thus being lack of electric moments. But, in the in-plane $x$ direction, the electric moments for M-$Bi_2O_2Se$ and M-$Bi_2O_2Te$ can also be built with a sizable value ~ 9.2 eÅ and ~ 9.5 eÅ. We can see that the in-plane spontaneous polarizations for Se and Te compounds are with similar strength since they are both mainly resulted from the asymmetric charge transfer in the top Bi-O parallelogram channel rather than the slightly deformed Janus Se(Te)-Bi-O layer lying below. The piezoelectric effect was additionally studied by applying bi-axial strain ranging from -2% to +2% inside the $x$-$y$ plane. And the dependence of vertical dipole moment on the strain was derived, as shown in Figure 4b. The polarization strength varies linearly with the in-plane deformation, which can be viewed as the solid evidence on spontaneous polarization in M-$Bi_2O_2Se(Te)$. The comparison between them also indicates that the piezoelectric effect for M-$Bi_2O_2Se$ is more remarkable than that for M-$Bi_2O_2Te$, in agreement with the magnitude of the vertical polarization value.

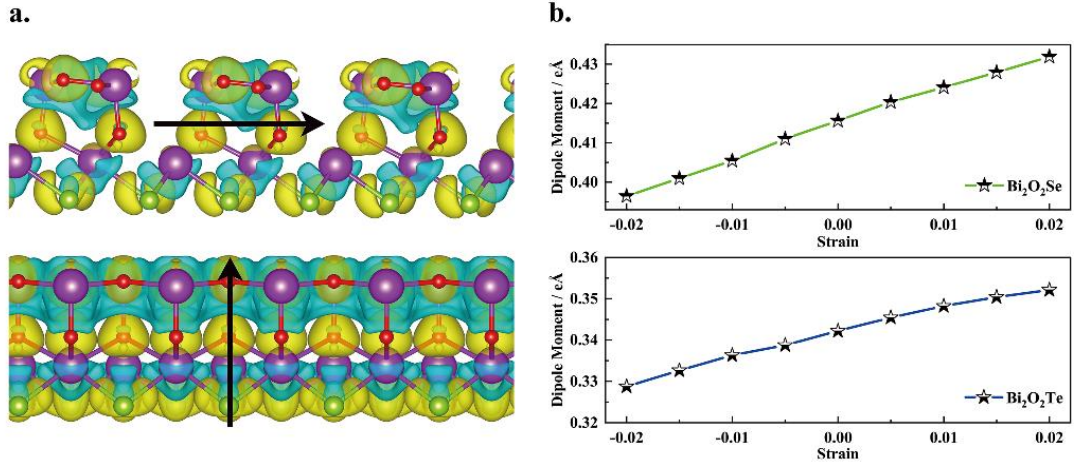

**Figure 4.** a. Side views of differential charge density for M-$Bi_2O_2Se$, where the iso-surface value is set to 0.005 e/Å$^3$. The spontaneous electric polarization in *x* and *z* directions are presented as black horizontal and vertical arrows; b. Calculated piezoelectric properties for M-$Bi_2O_2Se$ and M-$Bi_2O_2Te$, respectively.

## Conclusion

In conclusion, we theoretically predicted the stable monolayer phase for both $Bi_2O_2Se$ and $Bi_2O_2Te$ based on structure prediction algorithm. The simulation results clearly demonstrate that the conventional monolayer phase derived from bulk counterpart is not energetically favored and dynamically stable, and the reconstructed structure with strong in-plane and vertical anisotropy was revealed. Further calculation on electronic structures shows that the M-$Bi_2O_2Se$(Te) persists the ultra-high carrier mobility, especially for the electrons in M-$Bi_2O_2Te$ which can reach as high as 3610 cm$^2$V$^{-1}$s$^{-1}$ at room temperature. At the same time, moderate band gaps endow them excellent in-plane optical absorption property, which is even better than intrinsic silicon. While the remarkable spontaneous vertical electric polarization can also be realized because of such anisotropic reconstruction. This study not only clarified the monolayer phase of $Bi_2O_2Se$ with strong interlayer electrostatic interaction, but also opened up the great opportunity for their potential applications as the photo-electronic devices and field-effect transistors.

## Acknowledgements

Q.F.Z. was supported by National Key Research and Development Program of China (No. 2017YFB0702100) and National Natural Science Foundation of China (11404017). D.L. was supported by the European Regional Development Fund in the IT4Innovations national supercomputing center - path to exascale project, project number CZ.02.1.01/0.0/0.0/16_013/0001791 within the Operational Programme Research, Development and Education and grant No. 17-23964S of the Czech Science Foundations.

## Conflict of Interest

The authors declare no conflict of interest.

## Keywords

$Bi_2O_2Se$ monolayer; Structure prediction; High carrier mobility; Giant vertical electric polarization

Supporting Information

# Anisotropic $Bi_2O_2Se(Te)$ Monolayer: Realizing Ultra-High Carrier Mobility and Giant Electric Polarization in Two-Dimension

*Jiewen Xiao[†, §], Yuxin Wang[†, §], Xiang Feng[†, §], Dominik Legut[‡], Tianshuai Wang[†], Yanchen Fan[†], Tonghui Su[†], Rushi Gong[†], Ruifeng Zhang[†], Qianfan Zhang\*[†]*

[†]*School of Materials Science and Engineering, Beihang University, Beijing 100191, P. R. China.*

[‡]*IT4Innovations & Nanotechnology Centre, VSB-Technical University of Ostrava, 17.listopadu 15, Ostrava CZ-70833, Czech Republic.*

[§]**Author contributions:** *These authors contributed equally to this work.*

**\*Corresponding authors:** *qianfan@buaa.edu.cn.*

# Section I. Structure and Stability

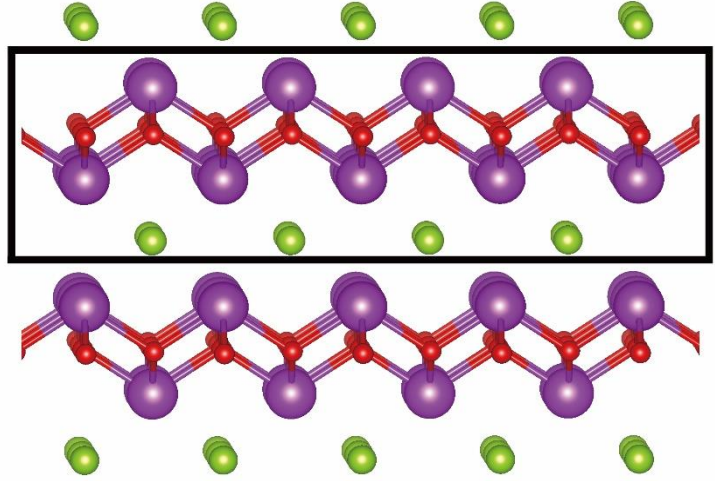

**Figure S1.** Layered bulk structure of $Bi_2O_2Se$ and the hypothesized monolayer phase is denoted in the black frame.

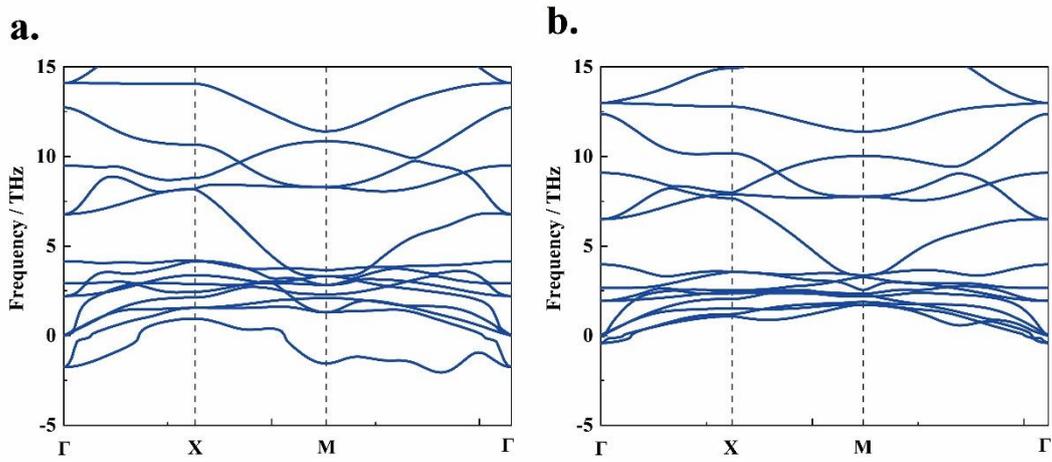

**Figure S2.** Phonon spectra of hypothesized monolayer for a. $Bi_2O_2Se$ and b. $Bi_2O_2Te$, respectively.

|  | a | b | c | α | β | γ |
|---|---|---|---|---|---|---|
| M-$Bi_2O_2Se$ | 6.792 | 4.271 | 26.00 | 90.00 | 90.00 | 90.00 |
| M-$Bi_2O_2Te$ | 7.070 | 4.337 | 26.00 | 90.00 | 90.00 | 90.00 |

**Table S1.** Lattice parameters of M-$Bi_2O_2Se$ and M-$Bi_2O_2Te$.

|      | **Fractional Coordinates** | | |
|:---:|:---:|:---:|:---:|
| **Atom** | *x* | *y* | *z* |
| Se | 0.010617 | 0.000000 | 0.397823 |
| Se | 0.508313 | 0.500000 | 0.398045 |
| O | 0.054062 | 0.000000 | 0.593688 |
| O | 0.700661 | 0.000000 | 0.506489 |
| O | 0.838791 | 0.500000 | 0.585241 |
| O | 0.191637 | 0.500000 | 0.506343 |
| Bi | 0.349675 | 0.000000 | 0.459315 |
| Bi | 0.734428 | 0.000000 | 0.586154 |
| Bi | 0.852744 | 0.500000 | 0.460839 |
| Bi | 0.159151 | 0.500000 | 0.586163 |

**Table S2.** Atomic positions of M-$Bi_2O_2Se$.

|      | **Fractional Coordinates** | | |
|:---:|:---:|:---:|:---:|
| **Atom** | *x* | *y* | *z* |
| Te | 0.011825 | 0.000000 | 0.392831 |
| Te | 0.509342 | 0.500000 | 0.393553 |
| O | 0.052452 | 0.000000 | 0.592475 |
| O | 0.707192 | 0.000000 | 0.507749 |
| O | 0.842498 | 0.500000 | 0.586738 |
| O | 0.180329 | 0.500000 | 0.506648 |
| Bi | 0.347462 | 0.000000 | 0.462088 |
| Bi | 0.745838 | 0.000000 | 0.587186 |
| Bi | 0.853956 | 0.500000 | 0.464264 |
| Bi | 0.149187 | 0.500000 | 0.586569 |

**Table S3.** Atomic positions of M-$Bi_2O_2Te$.

## Section II. Stability of the discovered M-Bi$_2$O$_2$Se(Te)

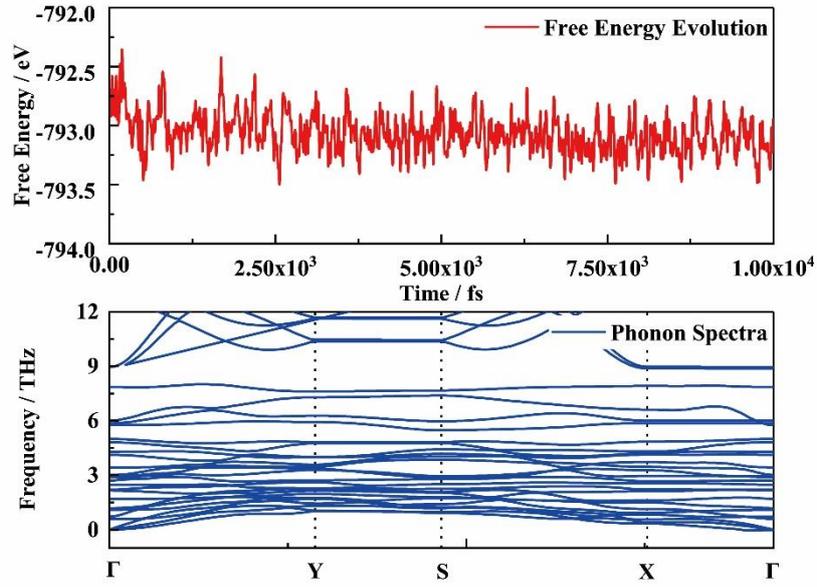

**Figure S4.** Free energy evolution during AIMD simulation (upper panel) and phonon spectra (lower panel) for M-Bi$_2$O$_2$Te.

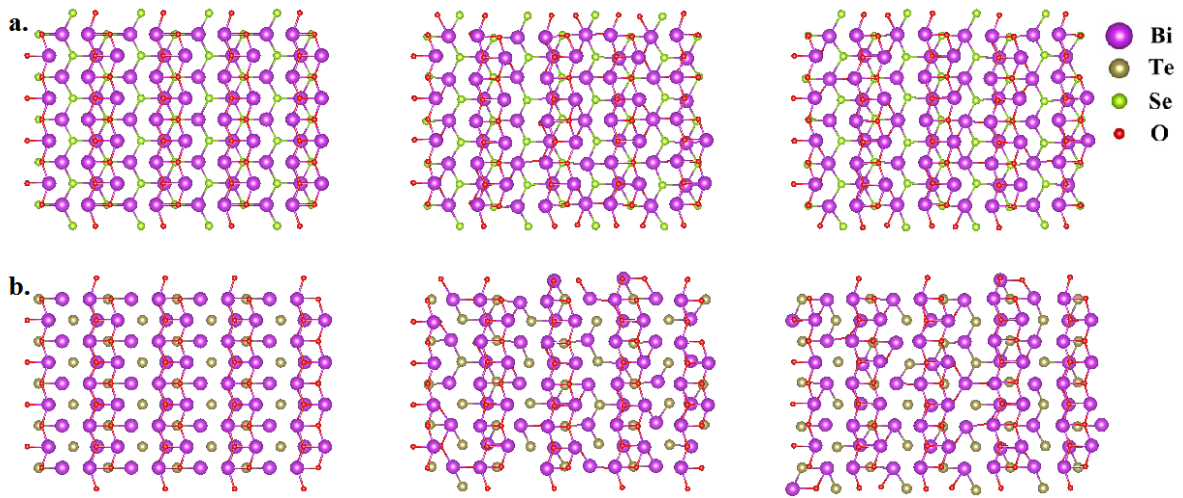

**Figure S5.** Atomic configurations for a. M-Bi$_2$O$_2$Se and b. M-Bi$_2$O$_2$Te after 0 *ps*, 5 *ps* and 10 *ps* AIMD simulation (from left to right).

## Section III. Electronic structures

As shown in Figure S6, the calculated band structure for bulk phase $Bi_2O_2Se$ exhibits an indirect band gap around 0.85 eV, being consistent with the value detected in experiments[36].

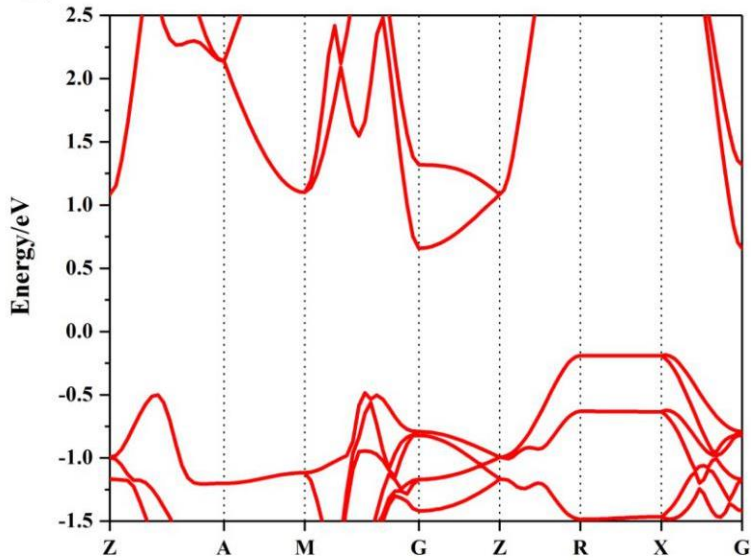

**Figure S6.** Calculated band structure of the bulk phase $Bi_2O_2Se$.

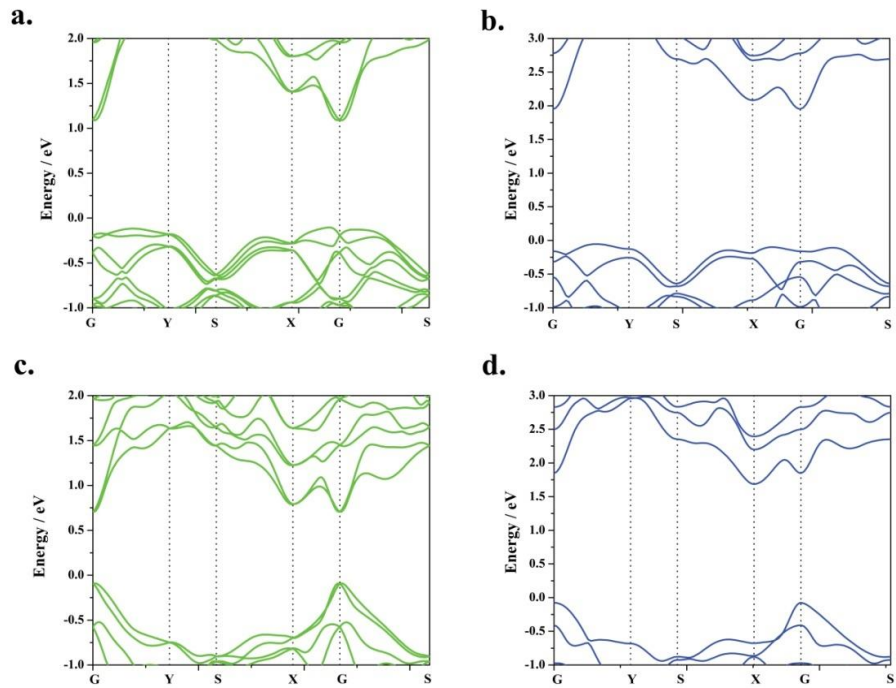

**Figure S7.** Calculated band structures based on a. PBE+SOC and b. HSE06 scheme for M-$Bi_2O_2Se$; Calculated band structures based on c. PBE+SOC and d. HSE06 scheme for M-$Bi_2O_2Te$.

## Section IV. Carrier Mobility Calculation

Another formulation for the phonon-limited scattering model, which is also widely used but can not accurately treat anisotropic systems[37]:

$$\mu_{\alpha x} = \frac{e\hbar C_{2D\_x}}{k_B T m_x^{*3/2} m_y^{*1/2} E_{\alpha x}^2}$$

All the parameters are with the same definition in the main text and calculated results are shown in table S4, being consistent with the first formulation.

|  | $C_x$ (N/m) | $C_y$ (N/m) | Carrier type | $m_x^*$ ($m_e$) | $m_y^*$ ($m_e$) | $E_{lx}$ (eV) | $E_{ly}$ (eV) | $\mu_x$ (cm$^2$V$^{-1}$s$^{-1}$) | $\mu_y$ (cm$^2$V$^{-1}$s$^{-1}$) |
|---|---|---|---|---|---|---|---|---|---|
| Bi$_2$O$_2$Se | 30.69 | 69.73 | e | 0.224 | 0.239 | 2.248 | 5.876 | 2499 | 779 |
|  |  |  | h | 0.153 | 1.107 | 2.268 | 3.912 | 2021 | 213 |
| Bi$_2$O$_2$Te | 27.03 | 70.75 | e | 0.136 | 0.127 | 3.144 | 5.092 | 3265 | 3489 |
|  |  |  | h | 0.221 | 0.228 | 2.680 | 2.734 | 1619 | 3947 |

**Table 4.** Calculated elastic modulus ($C_{2D\_x}$ and $C_{2D\_y}$), effective mass ($m_x^*$ and $m_y^*$), deformation potential ($E_x$ and $E_y$) and carrier mobility ($\mu_x$ and $\mu_y$) at 300 K.

## Supplementary References